# How Can The SN-GRB Time Delay Be Measured?

J. P. Norris and J. T. Bonnell

*Laboratory for High Energy Astrophysics,
NASA/Goddard Space Flight Center, Greenbelt, MD 20771*

**Abstract.** The connection between SNe and GRBs, launched by SN 1998bw / GRB 980425 and clinched by SN 2003dh / GRB 030329—with the two GRBs differing by a factor of ~ 50000 in luminosity—so far suggests a rough upper limit of ~ 1–2 days for the delay between SN and GRB. Only four SNe have had nonnegligible coverage in close coincidence with the initial explosion, near the UV shock breakout: two Type II, and two Type Ic, SN 1999ex and SN 1998bw. For the latter, only a hint of the minimum between the UV maximum and the radioactivity bump served to help constrain the interval between SN and GRB. Swift GRB alerts may provide the opportunity to study many SNe through the UV breakout phase: GRB 980425 "look alikes"—apparently nearby, low-luminosity, soft-spectrum, long-lag GRBs— accounted for half of BATSE bursts near threshold, and may dominate the Swift yield near threshold, since it has sensitivity to lower energies than did BATSE. The SN to GRB delay timescale should be better constrained by prompt UV/optical observations alerted by these bursts. Definitive delay measurements may be obtained if long-lag bursters are truly nearby: The SNe/GRBs could emit gravitational radiation detectable by LIGO-II if robust non-axisymmetric bar instabilities develop during core collapse, and/or neutrino emission may be detectable as suggested by Meszaros et al. [1].

## NATURE OF PROBLEM

The timescale between gamma-ray burst (GRB) onset and supernova collapse, $T_{0,GRB} - T_{0,SN} = \Delta T$, is characteristic of the progenitor's collapse. Knowledge of $\Delta T$ would constrain the physics relating GRB jet dynamics and SN core collapse. However, most GRBs with known redshifts are distant ($z_{median} \sim 1+$) and luminous, and thus SN onsets are swamped by the GRB afterglow. In fact without the GRB alert, the SN would remain undetected. The light curves of SNe are usually not well observed prior to the "radioactivity bump"—during the UV breakout phase and interjacent minimum. Since SN classification began in earnest in ~ 1992, more than 1700 SNe have been logged [2], but only four SNe have been observed for which any coverage of the shock breakout phase was obtained, two Type II(b) and two Type 1c (1987A, 1993J, 1998bw, and 1999ex [3–6]). The circumstances were unusual in each case. Both Type II's were bright; SN 1998bw was alerted by the GRB; and SN 1999ex was fortuitously discovered in the same galaxy being monitored for the progress of a Type II, SN 1999ee. Some details concerning the possible degree of constraint on $\Delta T$ derivable from observations of SNe onset are discussed in the next section. The prospects for constraints on $T_{0,SN}$ are of the order of ±0.5 day.

The best chance for empirical constraints on ΔT may come from long-lag GRBs whose distance scale is still ill-defined as a group. Long-lag GRBs may be relatively nearby compared to many luminous GRBs [7], and so as in the case of GRB 980425, detection and study of the SN onset may be much easier. But the Berger et al. [8] study of radio emission from Type Ic SNe suggests that most are less energetic than 1998bw, and may not be accompanied by GRBs. Thus important questions are, how frequently will we be able to study nearby SNe/GRBs, possibly like 1998bw/ 980425, and what channels will yield constraining information about ΔT and jet dynamics?

Modeling by Iwamoto et al. [9] yielded a predicted core collapse time for SN 1998bw within +0.7/–2 days of GRB 980425. In their Figure 1 illustrating light curves of three Type Ic SNe, the full-width half maxima span a dynamic range > 3. Thus, while SN 1998bw has often been quoted as a template, Type 1c SN light curves are highly variable one to the next, and seemingly unreliable for prophesying the extent of SN-like bumps in afterglows of distant GRBs, much less for predicting $T_{0,SN}$.

Predictions for ΔT vary over timescales of a few to ~ $10^6$ seconds, with timescales longer than 1–2 days excluded for one case so far (1998bw/980425 Iwamoto et al.). The central issue is whether the SN happens first (core-collapse) followed by the GRB (residual torus accretion), or if the events are (nearly) simultaneous, and this relates to strength of the two, or one, gravitational wave (GW) signals. Some current scenarios involve such a two-stage process with the timescale between collapse and explosion governed by accretion and other torques, and by black hole spin rate [10,11].

## BACKGROUND ON SUPERNOVA ONSETS

Observations of UV shock breakout in SNe are rarely obtained, due to a combination of factors. This phase is most prominent in the UV, whereas discovery proceeds from optical searches; the timescale from SN onset to end of breakout is short; and searches may require an increase in brightness (beginning of radioactive decay phase) to signal a SN and thence to proceed with observations to measure the light curve. Yet observing the rise of the UV breakout would best constrain $T_{0,SN}$ in the electromagnetic channel.

From examination of the light curves of the four SNe that have any observations of the onset and breakout phase, three trends are apparent: (1) Progressing through Type II (1987A), IIb (1993J), Ic (1999ex), and extreme Ic (1998bw), the hydrogen envelopes are increasingly depleted (then absent for Ic's), probably related to the duration of minima becoming shorter; this trend takes some study to discern due to the differing abscissa scales in the several references. (2) The minima tend to come earlier at redder colors (barely so for 1993J). (3) The minima are deeper with bluer color, to the extent that, in two of the cases, only an inflection is seen in R and redder colors. For 1998bw, coverage in (I, V, B, U) was less complete at early times than in V and R. Woosley et al. [12] discuss model UBV light curves for SN 1987A and Ib models, the first SN for which UV breakout was observed.

The closest in appearance of the four light curves are the two Type Ic, 1998bw and 1999ex. The pair have nearly equal radioactive decay widths, but different shapes especially near the minima, whose depths differ by two magnitudes. These shape variations near minimum emphasize the difficulty of using one Type 1c's light curve to infer an accurate onset time for another, even when the widths of the radioactive decay bumps are quite comparable.

The general conclusion must be that measuring SN onsets via the electromagnetic channel with sufficient accuracy required to constrain the interval between core collapse and GRB explosion to a small fraction of a day—comparable to some predictions [11]—may be at best difficult. Promising alternative approaches will probably still involve relatively nearby GRB sources and observations in neutrino or gravitational wave channels.

## LONG-LAG GAMMA-RAY BURSTS

The closest GRB sources may be those with relatively smooth emission, the individual pulse appearing as the canonical fast-rise exponential decay. These "long-lag" GRBs tend to have very few, wide pulses and spectral lags between BATSE energy bands of order 1 s or greater. This group dominates the BATSE sample of long bursts approaching trigger threshold, below a peak flux of ~ 0.7 photons cm$^{-2}$ s$^{-1}$. GRB 980425 was the famous example, for which the redshift is presumed to be z = 0.0085, associated with SN 1998bw. Whether a one or two-branch lag-luminosity relationship is more appropriate [13–15], may determine the average distance scale to these sources.

Several groups predicted subclasses similar to the observed long-lag bursts—with ultra-low luminosities compared to most GRBs, soft spectra, and possibly long spectral lags, where these properties may be attributed to a combination of low Lorentz factor, large jet opening angle, and/or large viewing angle [16–18]. However, the Berger et al. [8] survey of 32 Type 1c SNe found radio luminosities below that of 1998bw, suggesting that GRB 980425 may be anomalously and uniquely nearby. Regardless, long-lag bursts may be the closest subset given their tendency to occur near BATSE threshold, and thus may offer the best opportunities to explore ΔT in any channel, neutrino, gravitational or electromagnetic.

Swift's BAT should also detect many long-lag bursts. Their peak in νF(ν) clusters below 100 keV, compared to the median value for bright BATSE bursts of ~ 230 keV [19], and their lower power-law indices are steep. Calculations by Band [20] indicate that Swift's sensitivity to such bursts will be a factor of several better than BATSE's, as expected given the effective area vs. energy curve for Swift.

Fryer et al. [21] and Blondin et al. [22] discuss the possibility of bar instabilities developing during core collapse, giving rise to GWs. The ratio of rotational kinetic energy to gravitational potential must be sufficiently large. In an optimistic case [7], alerted by nearby long-lag GRBs, the GW emission would be detectable by LIGO II. Assuming 10 cycles, f ~ 200-800 Hz, source < 50 Mpc, then h/√Hz ~ 4 × 10$^{-24}$. This strain falls near the predicted LIGO threshold on the high frequency end. If the 2-branch lag-luminosity relationship obtained, then we would expect ~ 4 long-lag GRBs y$^{-1}$ (< 50 Mpc), with a rate within 100 Mpc ~ 30 yr$^{-1}$, with a few yr$^{-1}$ detectable by LIGO II. However, SN 1998bw may be anomalously nearby in this picture (its low luminosity partially attributable to viewing angle?), and most long-lag GRBs could be considerably more distant and undetectable in GWs.

## SUMMARY

From SN 1998bw/GRB 980425 and SN 2003dh/GRB 030329 we have conclusive evidence that highly energetic core-collapse SNe are associated with GRBs. To understand the SN/GRB relationship, we need to constrain $T_{0,GRB} - T_{0,SN} = \Delta T$ which is characteristic of the progenitor's collapse sequence. $\Delta T$ would also constrain the physics of GRB jet dynamics. For only four SNe, and only two Type Ic's, the UV shock breakout phase was observed. With such a small sample, and the inhomogeneity of Type Ic SNe light curves, it will be difficult to make progress. Moreover, GRB afterglows may be sufficiently bright to overwhelm the UV breakout rise to maximum, thus inhibiting investigation of timescales much shorter than currently probed for $\Delta T$. Regardless of their mean redshift (z ~ 0.01–0.1?), long-lag, soft-spectrum, low-brightness and thus apparently low-luminosity GRBs—possibly the closest subset of GRBs—may represent the best chance for constraining $\Delta T$. Long-lag bursts dominate near BATSE threshold, and their spectra are commensurate with Swift's BAT energy response. Thus we should detect a higher fraction of long-lag bursts with Swift. Eventually, the best channels for probing short $\Delta T$ s may be gravitational waves or neutrino emission [1]. However, depending on the mean distance to long-lag bursts and the particulars of GW and neutrino radiation, summation of several events over a timescale of a few years may be necessary to detect these signals.